# On Short-Range Enhancement of Van-der-Waals Forces

K. Makhnovets and A. Kolezhuk[*]

**Abstract**

We show that for nanoparticles with giant polarizability $\alpha$ (a prominent example being the recently studied $Na_{14}F_{13}$ molecular clusters) van der Waals forces are significantly enhanced at interparticle distances shorter than $R_0 \simeq (\alpha/2\pi\varepsilon_0)^{1/3}$. For an adequate description of this phenomenon, nonlinear effects must be taken into account. We show that, contrary to some theoretical claims, an accurate treatment of nonlinearity does not lead to any repulsive forces.

**Keywords:** dispersive interactions, giant polarizability, nonlinear polarizability, repulsive van der Waals interaction, mode softening

## 1. Introduction

In more than eight decades that passed since the seminal work by London [1], theory of dispersive, or Van der Waals (VdW) interactions has developed into a vast research field with numerous applications in physics, chemistry, and biology (see, e.g., [2-4] for a review). Despite the fact that the physics of VdW interactions is relatively well understood, this field continues to attract much attention [4-10], particularly due to the important role the VdW forces play at nanoscale, which is critical for various areas from catalysis [7] to molecular electronics [8] and self-assembly [9,10].

In this work, we focus on the VdW interaction of nanoparticles with the so-called "giant" polarizability. Under the latter term we understand the situation when the polarizability volume is significantly larger than the effective volume of the particle. Recently, giant polarizability, in excess of about 30 times the effective volume, has been observed in $Na_{14}F_{13}$ molecular clusters [11] and in some other alkali-halide clusters of the $M_nX_{n-1}$ type (see, e.g., [12] for a review). In those clusters, one electron occupies a loosely bound surface state outside a closed shell [13,14], and a simple quantum-mechanical model of a single electron moving on the surface of a sphere is capable to fit the experimental data on the temperature-dependent static polarizability [15,16]. At low temperatures, such clusters might exhibit the Jahn-Teller instability with a transition into a polar state; and a similar effect is also observed in niobium clusters [17,18].

[**] Corresponding author (e-mail: kolezhuk@gmail.com).
Institute of High Technologies, Taras Shevchenko National University of Kyiv,
Glushkova prosp. 4-G, 03022 Kyiv, Ukraine

In the present paper, we show that the VdW force between nanoparticles with anomalously large value of polarizability $\alpha$ is greatly enhanced when the interparticle distance $R$ becomes smaller than certain critical length $R_0 \simeq (\alpha/2\pi\varepsilon_0)^{1/3}$, where $\varepsilon_0$ is the vacuum permittivity. The essential physics of this effect can be understood already in the simplest framework of two isotropic point-like particles used in the original 1937 London's paper [1]: large polarizability causes softening of one of the dipole oscillation modes at $R = R_0$, with the mode energy vanishing as a square root of $R - R_0$, which leads to a divergence in the Van der Waals force. However, it is clear that such a picture is too simplistic: close to the mode softening point, nonlinear polarizability becomes important and has to be included into consideration, to render the system stable at $R < R_0$.

There are several theoretical works [19-21], studying a similar scenario of mode softening in presence of nonlinearity, for various setups of physisorption (point-like or ellipsoidal particles interacting with a dielectric or metallic surface, with and without surface plasmon waves). The above works came to the conclusion that nonlinearity causes the emergence of a strong repulsive contribution to the VdW force at short distances, leading to the Lennard-Jones type interaction potential. In the present paper, we show that a careful treatment of nonlinearity does not yield any contributions of the repulsive type, but merely leads to a removal of the unphysical divergence in the VdW force: instead of becoming infinite at $R = R_0$, the force is just strongly amplified by a factor inversely proportional to the nonlinear polarizability.

## 2. Model and analysis

Consider the system of two identical isotropic point-like polarizable particles in free space, separated by the distance $R$, described by the following Hamiltonian:

$$\hat{H} = \sum_{j=1,2} \left( \frac{\hat{\boldsymbol{\pi}}_{p,j}^2}{2m} + \frac{m\omega_0^2 \hat{\mathbf{p}}_j^2}{2} + \frac{\gamma \hat{\mathbf{p}}_j^4}{2} \right) + \frac{1}{4\pi\varepsilon_0 R^3} \left\{ (\hat{\mathbf{p}}_1 \cdot \hat{\mathbf{p}}_2) - 3(\hat{\mathbf{p}}_1 \cdot \mathbf{n})(\hat{\mathbf{p}}_2 \cdot \mathbf{n}) \right\}. \qquad (1)$$

Here $\hat{\mathbf{p}}_1$ and $\hat{\mathbf{p}}_2$ are the fluctuating dipole moments of the two particles, $\hat{\boldsymbol{\pi}}_1$ and $\hat{\boldsymbol{\pi}}_2$ are the corresponding canonically conjugate momenta, $\mathbf{n} = \mathbf{R}/R$ is the unit vector along the line connecting the particles, $\omega_0$ is the characteristic frequency of the dipole oscillations, the "effective mass" $m = (\alpha \omega_0^2)^{-1}$ is determined by the linear polarizability of the particle $\alpha$, and, finally, the nonlinear coupling $\gamma = |\beta|/(2\alpha^4)$ is related to the 3-rd order nonlinear polarizability $\beta$. We assume that the response of the dipole moment of a particle to the local field $\mathbf{E}$ is given by $\mathbf{p} = \alpha \mathbf{E} + \beta E^2 \mathbf{E} + \cdots$; for the sake of simplicity it is further assumed that $\beta < 0$ to ensure stability (otherwise one has to include higher-order nonlinear terms), and that particles have an inversion center so the 2-nd order nonlinear polarizability vanishes.

It is easy to see that the classical minimum of the energy determined by Eq. (1) is reached for $\mathbf{p}_1 = \mathbf{p}_2 = \pm p_0 \mathbf{n}$, with

$$p_0^2 = \begin{cases} 0, & R > R_0, \quad \text{where } R_0 \equiv (\alpha/2\pi\varepsilon_0)^{1/3} \\ p_m^2 \left( \dfrac{R_0^3}{R^3} - 1 \right), & R < R_0, \quad \text{where} \quad p_m^2 = \alpha^3/|\beta| \end{cases} \quad (2)$$

The above expression can be trusted only as long as $p_0$ remains small compared to the characteristic value $p_m$: obviously, for $p \simeq p_m$ the linear and nonlinear contributions to the dipole moment become of the same order of magnitude, so one has to include higher-order nonlinear terms.

In the spirit of the original London's paper, one can consider small harmonic fluctuations around the classical solution, setting $\mathbf{p}_{1,2} = \pm p_0 \mathbf{n} + \mathbf{u}_{1,2}$ and taking into account only quadratic terms in the fluctuations $\mathbf{u}_{1,2}$. Using the symmetric and antisymmetric normal modes $\mathbf{u}_\pm = (\mathbf{u}_1 \pm \mathbf{u}_2)/\sqrt{2}$, and passing to the "second quantization" form, one obtains the following Hamiltonian in the harmonic approximation:

$$\hat{H}_h = U_0 + \sum_{\sigma \in \{x\pm, y\pm, z\pm\}} \hbar \omega_\sigma \left( \hat{a}_\sigma^+ \hat{a}_\sigma + \frac{1}{2} \right), \quad (3)$$

where $U_0 = -\gamma p_0^4$. Here we have chosen the $z$ axis along the line connecting the particles, and the creation and annihilation operators $\hat{a}_\sigma^+$, $\hat{a}_\sigma$ are connected to the original operators of fluctuations $\mathbf{u}_\pm$ and their canonically conjugate momenta $\boldsymbol{\pi}_\pm$ via standard relations

$$u_\sigma = \left(\frac{\hbar}{2m\omega_\sigma}\right)^{1/2}(\hat{a}_\sigma^+ + \hat{a}_\sigma), \quad \pi_\sigma = i\left(\frac{m\hbar\omega_\sigma}{2}\right)^{1/2}(\hat{a}_\sigma^+ - \hat{a}_\sigma). \tag{4}$$

The normal mode frequencies $\omega_\sigma$ are for $R > R_0$ given by

$$\omega_{z\pm}^2 = \omega_0^2(1 \mp s), \quad \omega_{x\pm}^2 = \omega_{y\pm}^2 = \omega_0^2(1 \pm s/2), \quad \text{where} \quad s \equiv (R_0/R)^3, \tag{5}$$

and for $R < R_0$ by

$$\omega_{z+}^2 = 2\omega_0^2(s-1), \quad \omega_{z-}^2 = \omega_0^2(4s-2), \quad \omega_{x+}^2 = \omega_{y+}^2 = \frac{3\omega_0^2 s}{2}, \quad \omega_{x-}^2 = \omega_{y-}^2 = \frac{\omega_0^2 s}{2}. \tag{6}$$

The resulting ground state energy

$$U_h = U_0 + \frac{1}{2}\sum_\sigma \hbar\omega_\sigma \tag{7}$$

depends on the interparticle distance $R$. For large distances $R \gg R_0$ one recovers the well-known London's result [1]

$$U_h \approx U_L = \text{const} - \frac{3}{16}\hbar\omega_0\left(\frac{R_0}{R}\right)^6, \tag{8}$$

while at $R$ close to $R_0$, due to the softening of $(z+)$ mode, the energy has a square-root singularity $U_h \approx c_1 + c_2|R - R_0|^{1/2}$ and thus the Van der Waals force $F = -\partial U_h/\partial R$ diverges at $R \to R_0$. It is clear, however, that this "naïve" calculation neglecting anharmonicity cannot be trusted close to the softening point, where fluctuations of the dipole momenta become large, and nonlinear terms play crucial role. At $R$ slightly less then $R_0$, tunneling processes between two energetically equivalent states with $\mathbf{p}_{1,2} = \pm p_0 \mathbf{n}$ are dominant, and they are again determined by the nonlinear terms.

To improve the above theory at $R$ close to $R_0$, we will retain the full nonlinear description for $(z+)$ mode, which becomes "soft" ("slow") in this region, while nonlinear effects for the other five modes, which remain "hard" ("fast"), may be safely treated perturbatively. We again introduce symmetric and antisymmetric variables:

$$\mathbf{p}_\pm = (\mathbf{p}_1 \pm \mathbf{p}_2)/\sqrt{2}, \tag{9}$$

and cast the Hamiltonian in the following form:

$$\hat{H} = \hat{H}_{z+} + \sum_{\nu \in \{x\pm, y\pm, z-\}} \hat{H}_\nu + \hat{H}_{int}^{(1)} + \hat{H}_{int}^{(2)} + \hat{H}_{int}^{(3)}. \quad (10)$$

Here

$$\hat{H}_{z+} = \frac{\hat{\pi}_{z+}^2}{2m} + \frac{m\omega_0^2}{2} g_0 \hat{p}_{z+}^2 + \frac{1}{4}\gamma \hat{p}_{z+}^4, \quad g_0 = 1-s \quad (11)$$

corresponds to the nonlinear oscillator describing the "slow" mode, Hamiltonians $\hat{H}_\nu$ describe the harmonic part of five "fast" modes, and $\hat{H}_{int}^{(1)}$ is the part of interaction between "slow" and "fast" modes that is quadratic in $\hat{p}_{z+}$

$$\hat{H}_\nu = \hbar\omega_\nu \left( \hat{a}_\nu^+ \hat{a}_\nu + \frac{1}{2} \right),$$

$$\hat{H}_{int}^{(1)} = \frac{1}{2}\gamma \hat{p}_{z+}^2 \left( \hat{p}_{x+}^2 + \hat{p}_{x-}^2 + \hat{p}_{y+}^2 + \hat{p}_{y-}^2 + 3\hat{p}_{z-}^2 \right), \quad (12)$$

where frequencies $\omega_\nu$ are given by Eq. (5), and $\hat{p}_\nu$ coincide with $\hat{u}_\nu$, so their connection to the creation/annihilation operators is given by the same Eq. (4). Further, $\hat{H}_{int}^{(2)}$ describes the interaction between "fast" modes only,

$$\hat{H}_{int}^{(2)} = \frac{1}{4}\gamma \left( \hat{p}_{x+}^2 + \hat{p}_{x-}^2 + \hat{p}_{y+}^2 + \hat{p}_{y-}^2 + \hat{p}_{z-}^2 \right)^2 + \gamma \left( \hat{p}_{x+} \hat{p}_{x-} + \hat{p}_{y+} \hat{p}_{y-} \right)^2 \quad (13)$$

and, finally, $\hat{H}_{int}^{(3)}$ contains interaction terms linear in $\hat{p}_{z+}$:

$$\hat{H}_{int}^{(3)} = 2\gamma \hat{p}_{z-} \hat{p}_{z+} \left( \hat{p}_{x+} \hat{p}_{x-} + \hat{p}_{y+} \hat{p}_{y-} \right). \quad (14)$$

Now we can proceed to the analysis of the ground state energy for different regimes dictated by the interparticle distance. There are three such regimes determined by the interplay of two dimensionless parameters: quadratic coupling constant $g_0 = 1 - s \equiv 1 - (R_0 / R)^3$ (see Eqs. (5), (11)) and the quartic coupling $\lambda$ defined as

$$\lambda = \frac{\hbar\gamma}{m^2 \omega_0^3} = \frac{\hbar\omega_0}{2(\alpha^2 / |\beta|)}. \quad (15)$$

Physically, $\lambda$ is the ratio of the average zero-point fluctuation of the dipole moment of a single particle in the ground state (when the other particle is far away, $R \gg R_0$) to the characteristic "maximal" value $p_m$ (see Eq. (2)) beyond which the contribution of nonlinear terms to the energy starts to prevail over the linear ones. Hereafter we assume that $\lambda \ll 1$.

(a) *Weakly nonlinear (single-well) regime* $(1-s \gg \lambda^{2/3})$: in this case, the principal contribution to the ground state energy comes from harmonic terms and thus is given by the "naïve" expression Eq. (7). The potential for the dipole moment has a single-well type. All fluctuation modes are "hard", so nonlinear interactions can be taken into account perturbatively. In the first order in $\lambda$, the ground state energy (7) obtains corrections only from $\hat{H}_{int}^{(1)}$ and $\hat{H}_{int}^{(2)}$, and the result takes the form

$$U_a = \frac{\hbar \omega_0}{2}\left\{\sqrt{1-s} + \sqrt{1+s} + 2\sqrt{1-s/2} + 2\sqrt{1+s/2} + \lambda(f_1(s) + f_2(s))\right\}. \quad (16)$$

Here the corrections are separated into two parts $f_1$ and $f_2$ in such a way that $f_1$ is formally divergent at $s \to 1$ (i.e., at $R \to R_0$) and $f_2$ is not:

$$f_1(s) = \frac{3}{8}\left(\frac{\omega_0}{\omega_{z+}}\right)^2 + \frac{\omega_0}{2\omega_{z+}}\left(\frac{\omega_0}{\omega_{x+}} + \frac{\omega_0}{\omega_{x-}} + \frac{3\omega_0}{2\omega_{z-}}\right),$$

$$f_2(s) = \left(\frac{\omega_0}{\omega_{x+}} + \frac{\omega_0}{\omega_{x-}}\right)^2 + \frac{\omega_0}{2\omega_{z-}}\left(\frac{\omega_0}{\omega_{x+}} + \frac{\omega_0}{\omega_{x-}} + \frac{3\omega_0}{4\omega_{z-}}\right).$$
(17)

In this regime, the frequencies in Eq. (17) should be taken from Eq. (5).

(b) *Strongly nonlinear regime* $(|1-s| \ll \lambda^{2/3})$ corresponds to the vicinity of the mode softening point $R = R_0$. In this regime, for the "slow" mode $(z+)$ perturbation theory in $\lambda$ is not valid any more, and one should rather use the strong-coupling expansion [22] which is a series in the parameter $g_0(4/\lambda)^{2/3}$. On the typical time scale of oscillations of the "fast" $(\nu)$ modes, the average value of $\hat{p}_{z+}$ remains practically unchanged. For that reason, one can take into account the effects of interaction between the "slow" and "fast" modes by employing the procedure of averaging over "fast" modes and obtaining a renormalized effective potential for the slow mode. Such procedure is familiar in quantum field theory, where it is commonly known as "integrating out massive degrees of freedom". Doing so, one can see that in the leading order in $\lambda$, the effect of $\hat{H}_{int}^{(1)}$ amounts to the following renormalization of the quadratic coupling constant $g_0$ in the slow-mode Hamiltonian (11):

$$g_0 \mapsto g = 1 - s + \lambda \left( \frac{1}{\sqrt{1+s/2}} + \frac{1}{\sqrt{1-s/2}} + \frac{3}{2\sqrt{1+s}} \right), \qquad (18)$$

while the contributions from $\hat{H}_{int}^{(2)}$ merely shift the ground state energy, $U \mapsto U + \frac{1}{2}\hbar\omega_0 \lambda \cdot f_2(s)$, where $f_2(s)$ is defined as in Eq. (17), with frequencies taken from Eq. (5). The contribution from $\hat{H}_{int}^{(3)}$ vanishes in the first order in $\lambda$, and in the second order it yields an additional renormalization of $g$ proportional to $\lambda^2$, so it can be safely neglected. Summing up all those contributions, we obtain the ground state energy in this approximation:

$$\begin{aligned}U_b &= \hbar\omega_0 (\lambda/4)^{1/3} \sum_{n=0}^{\infty} c_n \left( (4/\lambda)^{2/3} g \right)^n \\ &+ \frac{\hbar\omega_0}{2} \left( \sqrt{1+s} + 2\sqrt{1-s/2} + 2\sqrt{1+s/2} + \lambda f_2(s) \right),\end{aligned} \qquad (19)$$

where the renormalized coupling $g$ is given by Eq. (18), and $c_n$ are the coefficients of the strong coupling perturbation expansion ($c_0 \approx 0.668$, $c_1 \approx 0.1437$, $c_2 \approx -0.0086$, …) that have been found by Janke and Kleinert [22] up to $n = 22$.

(c) *Weakly nonlinear double-well regime* $(s - 1 \gg \lambda^{2/3})$ corresponds to the situation of a double-well potential for the dipole moment, with "deep" wells. In this regime, the average magnitude of zero-point fluctuations of the dipole moment inside a well, roughly equal to $(\hbar/m\omega_{z+})^{1/2}$, is much smaller than the classical equilibrium magnitude of the dipole moment $p_0$, which determines the distance between the wells, so one can neglect the tunneling effects that will be exponentially small. All modes are "hard", so nonlinearity can again be treated by means of the perturbation theory in $\lambda$. As a result, one obtains for the ground energy

$$U_c \approx \frac{\hbar\omega_0}{2} \left\{ -\frac{(s-1)^2}{2\lambda} + \sqrt{2(s-1)} + \sqrt{4s-2} + \sqrt{2s} + \sqrt{6s} + \lambda \left( f_1(s) + f_2(s) \right) \right\}, \qquad (20)$$

where $f_1$, $f_2$ are given by Eq. (17), but in this case with the frequencies taken from (6). It should be remarked that since the condition $s - 1 \ll 1$ must still hold (in order for the dipole moment to remain small compared to $p_m$, so that our theory stays within its applicability range), the "window" of distances $R$ for this regime is rather narrow.

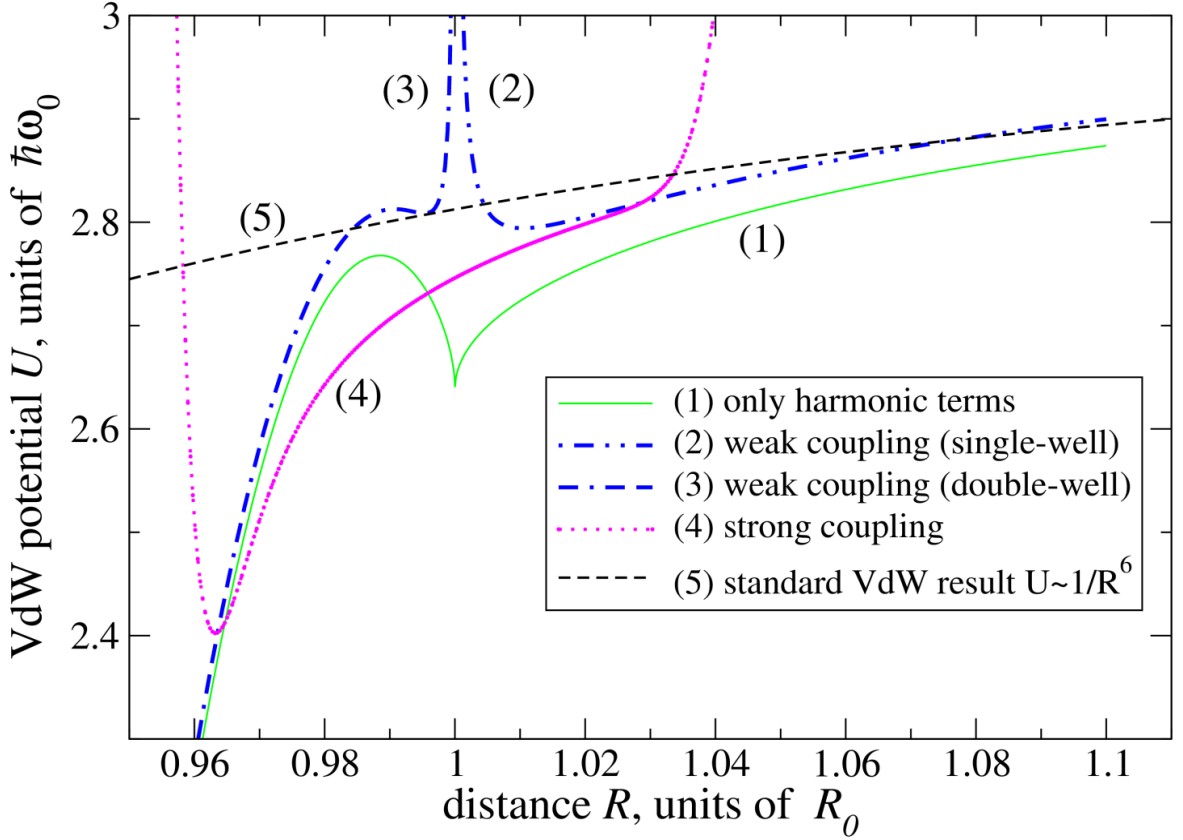

**Fig. 1.** The potential of Van der Waals interaction between two point-like particles separated by distance $R$, for the model described by Eq. (1), at the value $\lambda = 0.005$ of the quartic coupling parameter (15), obtained in various approximations. The dashed line 5 indicates the standard result (8) valid at $R \gg R_0$, solid line *1* corresponds to the harmonic approximation (7), dash-dotted lines *2,3* show the results obtained within the weak coupling perturbation theory for $R > R_0$ (Eq. (16)) and $R < R_0$ (Eq. (20)), respectively, and the dotted line *4* shows the result (19) obtained by means of the strong-coupling perturbation theory with the renormalized quadratic coupling (18).

## 3. Discussion

In order to better see the general picture and to understand the limitations of different approximations described in the previous section, it is instructive to look at Fig. 1, which shows the resulting Van der Waals interaction potential for the value $\lambda = 0.005$ of the nonlinear coupling constant (15).

One can see that for the weak nonlinearity ($\lambda \ll 1$) the results of the "naïve" harmonic approximation (7), shown by the solid line in Fig. 1, describe the VdW potential relatively well but become inadequate in the vicinity of the critical point $R = R_0$, as expected. The usual weak-coupling perturbation theory leads merely to small corrections away from the critical point, but fails badly in its vicinity; the corresponding results, given by Eq. (16) for $R > R_0$ and Eq. (20) for $R < R_0$, are shown in Fig. 1 by dash-dotted lines. The reason for this failure is obviously the softening of $(z+)$ mode: since its frequency vanishes at $R = R_0$, the corresponding correction $f_1(s)$ diverges (see Eq.(17)). In contrast to that, our result (19), based on the strong-coupling perturbation theory for the soft mode, taking into account the interaction between the soft mode and "hard" modes via the renormalization of the quadratic coupling (18), and combined with the weak-coupling corrections from "hard" modes, is able to capture correctly the behavior of the VdW interaction in the vicinity of the critical point $R = R_0$ (see the dotted line in Fig. 1).

The standard $1/R^6$ result (8), shown with the dashed line in Fig. 1, remains valid at $R \gg R_0$, but strongly underestimates the VdW force at distances about $R_0$ and less. One can roughly define the "enhancement factor" $\eta$ as the ratio of the actual force at $R = R_0$ (calculated with the help of our formula (19)) to the expression which follows from London's result Eq. (8):

$$\eta = \frac{(\partial U_b / \partial R)}{(\partial U_L / \partial R)}\bigg|_{R=R_0} \sim \frac{1}{\lambda^{1/3}}, \qquad (21)$$

so for typical values of $\lambda \sim 10^{-3}$ [23] one can expect an enhancement of about one order of magnitude.

Another observation one can make from our analysis is that a careless use of the weak-coupling perturbation theory might be dangerous. Indeed, even though a small value of $\lambda \ll 1$ means that for a single particle the effects of nonlinear polarization are small and thus can be treated perturbatively in $\lambda$, in the situation when one of the modes is driven soft by some interaction, the frequency $\omega_0$ in the definition (15) of the coupling $\lambda$ should be replaced by the frequency of the soft mode (in our case $\omega_0(1-s)^{1/2}$), so that the actual small parameter for the soft mode is not $\lambda$ but rather $\lambda/(1-s)^{3/2}$. Looking at the weak-coupling perturbative result Eq. (16) for $R > R_0$, and at the corresponding dash-dotted curve in Fig. 1, one can see that a careless extrapolation of the weak-coupling perturbative expression, valid for large distances, on the regime of strong coupling leads to a spurious repulsion at short distances close to the mode softening point. Such an artifact strongly resembles the nonlinearity-induced repulsive VdW interaction claimed in Refs. [19-21] on the basis of a weak-coupling perturbative treatment.

## 4. Summary

In the framework of a simple model describing two isotropic nanoparticles with a giant linear polarizability $\alpha$ (i.e., with the polarizability volume $\alpha' = \alpha/(4\pi\varepsilon_0)$ substantially larger than the particle volume) and stabilizing third-order nonlinear polarizability $\beta < 0$, we have shown that Van der Waals forces are significantly enhanced at interparticle distances shorter than $R_0 \simeq (2\alpha')^{1/3}$. The enhancement factor is roughly given by $\left(\dfrac{\hbar\omega_0}{\alpha^2/|\beta|}\right)^{-1/3}$, where $\omega_0$ is the characteristic frequency of the dipole oscillations. We have also shown that an accurate treatment of nonlinearity does not lead to any repulsive forces, contrary to some recent theoretical claims [19-21]. Our results may be important for theoretical understanding of Van der Waals interactions in systems of alkali-halide molecular clusters [11] with the anomalously large polarizability.

## Acknowledgements

We are grateful to V. Lozovski for fruitful discussions. This work was partly supported by grant 14БФ07-01 from the Ministry of Science and Education of Ukraine.

## References

[1] R. Eisenschitz and F. London, *Z. Phys.* **1930,** *60,* 491; F. London, *Trans. Faraday Soc.,* **1937**, *33*, 8.

[2] Yu.S.Barash, *Van der Waals Forces,* Nauka, Moscow **1988**.

[3] V. Adrian Parsegian, *Van Der Waals Forces:A Handbook For Biologists, Chemists, Engineers, And Physicists*, Cambridge University Press, Cambridge, New York **2006.**

[4] D. Dalvit, P. Milonni, D. Roberts, F. da Rosa (Eds), *Casimir Physics* (Lecture Notes in Physics **834**, Springer, Berlin 2011), 457 pp.

[5] L. Novotny and C. Henkel, *Optics Letters* **2008**, *33*, 1029.

[6] S. V. Aradhya, M. Frei, M. S. Hybertsen, and L. Venkataraman, *Nature Materials,* **2012**, *11*, 872.

[7] J. K. Norskov, T. Bligaard, J. Rossmeisl, and C. H. Christensen, *Nature Chem.* **2009**, *1*, 37.

[8] K. Moth-Poulsen and T. Bjornholm, *Nature Nanotech.* **2009**, *4*, 551.

[9] L. Bartels, *Nature Chem.* **2010**, *2*, 87.

[10] G. Singh, H. Chan, A. Baskin, E. Gelman, N. Repnin, P. Král, and R. Klajn, *Science* **2014**, *345*, 1149.

[11] D. Rayane, I. Compagnon, R. Antoine, M. Broyer, Ph. Dugourd, P. Labastie, J. M. L'Hermite, A. Le Padellec, G. Durand, F.Calvo, F. Spiegelman, and A. R. Allouche, *J. Chem. Phys.* **2002**, *116*, 10730.

[12] K. Hoang, M.-S. Lee, S. D. Mahanti, and P. Jena, *Clusters: An Embryonic Form of Crystals and nanostructures.* In: P. Jena and A. W. Castleman (Eds), *Nanoclusters: a bridge across disiplines*, Elsevier, Amsterdam **2010,** Chapter 2, pp. 37-71.

[13] U. Landman, D. Scharf, and J. Jortner, *Phys. Rev. Lett.* **1985**, *54*, 1860.

[14] G. Durand, J. Giraud-Girard, D. Maynau, F. Spiegelmann, and F. Calvo, *J. Chem. Phys.* **1999**, *110*, 7871.

[15] G. Durand, F. Spiegelman, A. R. Allouche, *Eur. Phys. J. D* **2003,** *24,*19.

[16] A. V. Shytov and P. B. Allen, Phys. Rev. B **2006**, *74*, 075419.

[17] R. Moro, X. Xu, S. Yin, W. A. de Heer, *Science* **2003**, *300*, 1265.

[18] X. Xu, S. Yin, R. Moro, A. Liang, J. Bowlan, and W. A. de Heer, *Phys. Rev. B* **2007**, *75*, 085429.

[19] V.Z. Lozovskii and B.I. Khudik, *Phys. Stat. Sol. (b)* **1990**, *158*, 511; *ibid.*, **1990**, *160*, 137.

[20] V.Lozovski and V. Piatnytsia, *J. Comp. Theor. Nanoscience* **2013**, *10*, 2288.

[21] D. Kysylychyn, V. Piatnytsia, V. Lozovski, *Phys. Rev. E* **2013**, *88*, 052403.

[22] W. Janke and H. Kleinert, *Phys. Rev. Lett.* **1995**, *75*, 2787.

[23] R.W. Boyd, *Nonlinear Optics,* Academic Press, New York **2008**.